\documentclass[useAMS,usetimes,usemathptm,usenatbib,usegraphicx]{mn2e}

\title[Dynamic stabilization of non-spherical bodies against
unlimited collapse]{Dynamic stabilization of non-spherical
bodies\\
against unlimited collapse}
\author[G. S. Bisnovatyi-Kogan and O. Yu. Tsupko]{G. S. Bisnovatyi-Kogan$^{1,2,3}$\thanks{E-mail:
gkogan@iki.rssi.ru (GSBK); tsupko@iki.rssi.ru (OYuT)} and O. Yu.
Tsupko$^{1,3}$\footnotemark[1]\\
$^{1}$Space Research Institute of Russian Academy of Science, Profsoyuznaya 84/32, Moscow 117997, Russia\\
$^{2}$Joint Institute for Nuclear Research, Dubna, Russia\\
$^{3}$Moscow Engineering Physics Institute, Moscow, Russia}
\begin{document}

\pagerange{\pageref{firstpage}--\pageref{lastpage}} \pubyear{2007}

\maketitle

\label{firstpage}

\begin{abstract}
We solve equations, describing in a simplified way the newtonian
dynamics of a selfgravitating nonrotating spheroidal body after
loss of stability. We find that contraction to a singularity
happens only in a pure spherical collapse, and deviations from the
spherical symmetry stop the contraction by the stabilising action
of nonlinear nonspherical oscillations. A real collapse happens
after damping of the oscillations due to energy losses, shock wave
formation or viscosity. Detailed analysis of the nonlinear
oscillations is performed using a Poincar\'{e} map construction.
Regions of regular and chaotic oscillations are localized on this
map.
\end{abstract}


\section{Introduction}

Dynamic stability of spherical stars is determined by an average
adiabatic power $\gamma=\frac{\partial\log
P}{\partial\log\rho}|_S$. For a density distribution $\rho=\rho_0
\varphi(m/M)$, the star in the newtonian gravity is stable against
dynamical collapse when
$\int_0^R{(\gamma-\frac{4}{3})P\frac{dm}{\varphi(m/M)}}>0$
\citep{ZN, BK1989}. This approximate criterium becomes exact for
adiabatic stars with constant $\gamma$. Here $\rho_0$ is a central
density, $M$ is a stellar mass, $m$ is the mass inside a
Lagrangian radius $r$, so that $m=4\pi\int_0^r{\rho r^2 dr}$,
$M=m(R)$, $R$ is a stellar radius. Collapse of a spherical star
may be stopped only by a stiffening of the equation of state, like
neutron star formation at late stages of evolution, or formation
of fully ionized stellar core with $\gamma=\frac{5}{3}$ at
collapse of clouds during star formation. Without a stiffening a
spherical star in the newtonian theory would collapse into a point
with an infinite density (singularity).

In the presence of a rotation a star is becoming more dynamically
stable against collapse. Due to the more rapid increase of a
centrifugal force during contraction, in comparison with the
newtonian gravitational force, collapse of a rotating star will be
always stopped at finite density by centrifugal forces. Here we
show, that deviation from the spherical symmetry in a non-rotating
star with zero angular momentum leads to a similar stabilization,
and non-spherical star without dissipative processes  never will
reach a singularity. Therefore collapse to a singularity is
connected with a secular type of instability, even without
rotation.

When a uniform sphere of cold gas is set into free-fall collapse
under its self-gravity its shape is unstable. The linear
instability of large-scale modes (second-order harmonics) during
the collapse was discovered by \citet{LB64} and for the case of
non-rotating sphere by \citet{LB79} (see also \citet{LB96}).
Numerical investigations of collapsing pressureless spheroids have
been done in the papers of \citet{LB64} and \citet{LMS}. In
\citet{LB79} it was also shown that the pressure prevents the
development of large-scale shape instability if initially the
gravity is more than three-fifths pressure resisted. \citet{Fuji}
derived the equations of motion for a rotating ellipsoid from a
system of hydrodynamical equations. However, the accounting for
pressure effects in his approach was inconsistent with thermal
processes, leading to the wrong results for the dynamical
behaviour of a system with radiative losses (see
\citet{BKTs2005}). Using the variational principle and deriving
the equation for the entropy function from energy balance give the
correct relations for the pressure and the total energy. In
\citet{RT91} the virial equations for rotating Riemann ellipsoids
of incompressible fluid are demonstrated to form a Hamiltonian
dynamical system.

We calculate a dynamical behavior of a non-spherical, non-rotating
star after its loss of a linear stability, and investigate
nonlinear stages of contraction. We use approximate system of
dynamic equations, describing 3 degrees of freedom of a uniform
self-gravitating compressible ellipsoidal body (\citet{BK2004},
\citet {BKTs2005}). We obtain that the development of instability
leads to the formation of a regularly or chaotically oscillating
body, in which dynamical motion prevents the formation of the
singularity. We find regions of chaotic and regular pulsations by
constructing a Poincar\'e diagram for different values of the
initial eccentricity and initial entropy. For simplicity we
restrict ourself by calculating only spheroidal figures with
$\gamma=\frac{4}{3}$, and only briefly represent results for
$\gamma=\frac{6}{5}$. At the end we discuss qualitatively effects
of general relativity in a non-spherical collapse of a
non-rotating body.

\section{Equations of motion}

Let us consider 3-axis ellipsoid with semi-axes $a \neq b \neq c$:

\begin{equation}
\label{ellipsoid}\frac{x^2}{a^2}+\frac{y^2}{b^2}+\frac{z^2}{c^2}=1,
\end{equation}
and uniform density  $\rho$. A mass $m$ of the uniform ellipsoid
is written as ($V$ is the volume of the ellipsoid)

\begin{equation}
\label{mass}m = \rho \, V = \frac{4\pi}{3} \, \rho \, abc \, .
\end{equation}
Let us assume a linear dependence of velocities on coordinates

\begin{equation}
\label{linear-dep} \upsilon_x =\frac{\dot{a}x}{a} \: , \quad
\upsilon_y =\frac{\dot{b}y}{b} \: , \quad \upsilon_z
=\frac{\dot{c}z}{c} \, .
\end{equation}
The gravitational energy of the uniform ellipsoid is defined as
\citep{ll93}:

\begin{equation}
\label{grav-energy}
U_g=-\frac{3Gm^2}{10}\int\limits_0^{\infty}\frac{du}{\sqrt{(a^2+u)(b^2+u)(c^2+u)}}.
\end{equation}
The equation of state $P=K\rho^\gamma$ is considered here, with
$\gamma =4/3$. Note that the case $\gamma=5/3$ was considered by
\citet{BK2004} and \citet{BKTs2005}, but this case is not
interesting for the present work, because isentropic spherical star
with $\gamma=5/3$ always stops contraction, and never suffers
collapse to singularity. A spherical star with $\gamma =4/3$
collapses to singularity at small enough $K$, and we show here, how
deviations from a spherical form prevent formation of any
singularity. For $\gamma = 4/3$, the thermal energy of the ellipsoid
is $E_{th} \sim V^{-1/3} \sim (abc)^{-1/3}$, and the value
\[
\varepsilon=E_{th}(abc)^{1/3} =3\left(\frac{3m}{4\pi}\right)^{1/3}K
\]
remains constant in time. A Lagrange function of the ellipsoid is
written as
\begin{equation}
\label{Lagr-funct} L = U_{kin} - U_{pot}\, , \quad U_{pot} = U_g +
E_{th} \: ,
\end{equation}
\begin{equation}
\label{U-kin} U_{kin} = \frac{1}{2} \,
\rho\int\limits_V(\upsilon_x^2+\upsilon_y^2+\upsilon_z^2)\,dV
 = \frac{m}{10} \,
(\dot{a}^2+\dot{b}^2+\dot{c}^2) \: ,
\end{equation}
\begin{equation}
\label{E-th} E_{th} = \frac {\varepsilon} {(abc)^{1/3}} \: .
\end{equation}
Equations of motion describing behavior of 3 semiaxes $(a,b,c)$ is
obtained from the Lagrange function (\ref{Lagr-funct}) in the form

\begin{equation}
\ddot{a} = - \frac{3Gm}{2} \, a
\int\limits_0^{\infty}\frac{du}{(a^2+u)\Delta} \,\, +
\,\frac{5}{3m}\, \frac{1}{a} \,\frac {\varepsilon} {(abc)^{1/3}}
\, \, ,
\end{equation}
\begin{equation}
\ddot{b} = - \frac{3Gm}{2} \, b
\int\limits_0^{\infty}\frac{du}{(b^2+u)\Delta} \,\,+
\,\frac{5}{3m}\, \frac{1}{b} \,\frac {\varepsilon} {(abc)^{1/3}}
\, \, ,
\end{equation}
\begin{equation}
\ddot{c} = - \frac{3Gm}{2} \, c
\int\limits_0^{\infty}\frac{du}{(c^2+u)\Delta} \,\,+
\,\frac{5}{3m}\, \frac{1}{c} \,\frac {\varepsilon} {(abc)^{1/3}}
\; ,
\end{equation}
\[
\Delta^2 = (a^2+u)(b^2+u)(c^2+u) \, .
\]

\section{Dimensionless equations and numerical results}

To obtain a numerical solution of equations we write them in
non-dimensional variables. Let us introduce the variables
\[
\tilde{t} = \frac{t}{t_0}, \; \tilde{a} = \frac{a}{a_0}, \;
\tilde{b} = \frac{b}{a_0}, \; \tilde{c} = \frac{c}{a_0}, \;
\]
\[
\tilde{m} = \frac{m}{m_0}, \; \tilde{\rho} = \frac{\rho}{\rho_0},
\; \tilde{U} = \frac{U}{U_0}, \; \tilde{E}_{th} =
\frac{E_{th}}{U_0}, \; \tilde{\varepsilon} =
\frac{\varepsilon}{\varepsilon_0}.
\]
The scaling parameters $\, t_0, \; a_0, \; m_0, \; \rho_0, \; U_0,
\; \varepsilon_0$ are connected by the following relations
\begin{equation}
\label{scal-par} t_0^2 = \frac{a_0^3}{G m_0}, \, U_0 = \frac{G
m_0^2}{a_0}, \, \rho_0 = \frac{m_0}{a_0^3}, \, \varepsilon_0 = U_0
a_0.
\end{equation}
System of non-dimensional equations:
\begin{equation}
\label{eq28}
 \ddot{a} = - \frac{3m}{2} \, a
\int\limits_0^{\infty}\frac{du}{(a^2+u)\Delta} \,\, +
\,\frac{5}{3m}\, \frac{1}{a} \,\frac {\varepsilon} {(abc)^{1/3}}
\, \, ,
\end{equation}
\begin{equation}
\label{eq29} \ddot{b} = - \frac{3m}{2} \, b
\int\limits_0^{\infty}\frac{du}{(b^2+u)\Delta} \,\,+
\,\frac{5}{3m}\, \frac{1}{b} \,\frac {\varepsilon} {(abc)^{1/3}}
\, \, ,
\end{equation}
\begin{equation}
\label{eq30} \ddot{c} = - \frac{3m}{2} \, c
\int\limits_0^{\infty}\frac{du}{(c^2+u)\Delta} \,\,+
\,\frac{5}{3m}\, \frac{1}{c} \,\frac {\varepsilon} {(abc)^{1/3}}
\; ,
\end{equation}
\[
\Delta^2 = (a^2+u)(b^2+u)(c^2+u) \, .
\]

In equations (\ref{eq28})-(\ref{eq30}) only non-dimensional
variables are used, and "tilde" sign is omitted for simplicity in
this section. The non-dimensional Hamiltonian (or non-dimensional
total energy) is:
\[
H = U_{kin} + U_g + E_{th} = \frac{m}{10} \,
(\dot{a}^2+\dot{b}^2+\dot{c}^2) -
\]
\begin{equation}
\label{ham} -
\frac{3m^2}{10}\int\limits_0^{\infty}\frac{du}{\sqrt{(a^2+u)(b^2+u)(c^2+u)}}
+ \frac {\varepsilon} {(abc)^{1/3}} \, .
\end{equation}
In case of the sphere ($a=b=c$, $\dot{a}=\dot{b}=\dot{c}$) the
non-dimensional Hamiltonian and non-dimensional equations of
motion reduce to:
\begin{equation} \label{sphere-H}
H = \frac{3}{10} \,m \dot{a}^2 -\frac{3}{5a} \left( m^2 -
\frac{5}{3} \, {\varepsilon} \right) ,
\end{equation}
\begin{equation}
\label{sphere1-H} \ddot{a} = - \frac{1}{m a^2} \left( m^2 -
\frac{5}{3} \, \varepsilon \right) .
\end{equation}
As follows from (\ref{sphere-H}), (\ref{sphere1-H}) for the given
mass there is only one equilibrium value of $\varepsilon$
\begin{equation}
\label{sphere2-H}
 \varepsilon_{eq}=\frac{3m^2}{5},
\end{equation}
at which the spherical star has zero total energy, and it may have
an arbitrary radius. For smaller $\varepsilon<\varepsilon_{eq}$
the sphere should contract to singularity, and for
$\varepsilon>\varepsilon_{eq}$ there will be a total disruption of
the star with an expansion to infinity. We solve here numerically
the equations of motion for a spheroid with $a=b \neq c$, which,
using (\ref{eq28})-(\ref{eq30}), are written for the oblate
spheroid with $k=c/a<1$ as

\begin{equation}
\label{eq7} \ddot{a}=\frac{3}{2}\frac{m}{a^2(1-k^2)}
\biggl[k-\frac{\arccos{k}}{\sqrt{1-k^2}}\biggr]+\, \frac{5}{3m}\,
\frac{1}{a} \,\frac {\varepsilon} {(a^2c)^{1/3}},
\end{equation}

\begin{equation}
\label{eq8} \ddot{c}=-3\frac{m}{a^2(1-k^2)}
\biggl[1-\frac{k\arccos{k}}{\sqrt{1-k^2}}\biggr]+ \frac{5}{3m}\,
\frac{1}{c} \,\frac {\varepsilon} {(a^2c)^{1/3}};
\end{equation}
and for the prolate spheroid $k=c/a>1$ as

\begin{equation}
\label{eq7p} \ddot{a}=-\frac{3}{2}\frac{m}{a^2(k^2-1)}
\biggl[k-\frac{\cosh^{-1}{k}}{\sqrt{k^2-1}}\biggr]+ \frac{5}{3m}\,
\frac{1}{a} \,\frac {\varepsilon} {(a^2c)^{1/3}},
\end{equation}

\begin{equation}
\label{eq8p} \ddot{c}=3\frac{m}{a^2(k^2-1)}
\biggl[1-\frac{k\cosh^{-1}{k}}{\sqrt{k^2-1}}\biggr]+
\frac{5}{3m}\, \frac{1}{c} \,\frac {\varepsilon} {(a^2c)^{1/3}}.
\end{equation}
It is convenient to introduce variables

\begin{equation}
\label{eqvar}
 \varepsilon_*=
\frac{5}{3}\frac{\varepsilon}{m^2},\,\,\, t_*=t\sqrt m.
\end{equation}
In these variables the equations (\ref{eq7})-(\ref{eq8p}),
(\ref{sphere1-H}) are written as (omitting subscript "*")

\begin{equation}
\label{equ7} \ddot{a}=\frac{3}{2a^2(1-k^2)}
\biggl[k-\frac{\arccos{k}}{\sqrt{1-k^2}}\biggr]+\,
 \frac{1}{a} \,\frac {\varepsilon} {(a^2c)^{1/3}},
\end{equation}

\begin{equation}
\label{equ8} \ddot{c}=-\frac{3}{a^2(1-k^2)}
\biggl[1-\frac{k\arccos{k}}{\sqrt{1-k^2}}\biggr]+
 \frac{1}{c} \,\frac {\varepsilon} {(a^2c)^{1/3}}
\end{equation}
 for the oblate spheroid $k=c/a<1$,

\begin{equation}
\label{equ7p} \ddot{a}=-\frac{3}{2a^2(k^2-1)}
\biggl[k-\frac{\cosh^{-1}{k}}{\sqrt{k^2-1}}\biggr]+
 \frac{1}{a} \,\frac {\varepsilon} {(a^2c)^{1/3}},
\end{equation}

\begin{equation}
\label{equ8p} \ddot{c}=\frac{3}{a^2(k^2-1)}
\biggl[1-\frac{k\cosh^{-1}{k}}{\sqrt{k^2-1}}\biggr]+
 \frac{1}{c} \,\frac {\varepsilon} {(a^2c)^{1/3}}
\end{equation}
for the prolate spheroid $k=c/a>1$, and

\begin{equation}
\label{sphere3-H} \ddot{a} = - \frac{1 -  \varepsilon}{a^2}
\end{equation}
for the sphere, where the equilibrium corresponds to
$\varepsilon_{eq}=1$. Near the spherical shape we should use
expansions around $k=1$, what leads to equations of motion valid for
both oblate and prolate cases

\[
\ddot{a} = - \frac{1 -  \varepsilon}{a^2}+
\left(\frac{\varepsilon}{3}+\frac{3}{5}\right)\frac{1-k}{a^2},
\]
\begin{equation}
\label{sphere4-H} \ddot{c} = - \frac{1 -  \varepsilon}{a^2}+
\left(\frac{4\varepsilon}{3}-\frac{4}{5}\right)\frac{1-k}{a^2}.
\end{equation}
In these variables the total energy is written as

\begin{equation}
\label{eqvar1} H_*=\frac{H}{m^2},
\end{equation}
and omitting "*" we have

\[
H= \frac{\dot{a}^2}{5}+\frac{\dot{c}^2}{10}
-\frac{3}{5a}\frac{\arccos{k}}{\sqrt{1-k^2}} + \frac{3}{5}\frac
{\varepsilon} {(a^2c)^{1/3}},\quad {\rm (oblate)}
\]

\[
H= \frac{\dot{a}^2}{5}+\frac{\dot{c}^2}{10}
-\frac{3}{5a}\frac{\cosh^{-1}{k}}{\sqrt{k^2-1}} + \frac{3}{5}\frac
{\varepsilon} {(a^2c)^{1/3}}, \quad {\rm (prolate)}
\]
\begin{equation}
H = \frac{3\dot{a}^2}{10} -\frac{3}{5a} (1- {\varepsilon}), \quad
{\rm (sphere)} \label{ham1}
\end{equation}
\[
H= \frac{\dot{a}^2}{5}+\frac{\dot{c}^2}{10}
-\frac{3}{5a}\left(1+\frac{\delta}{3}+\frac{2\delta^2}{15}\right)
+\frac{3\varepsilon}{5a}\left(1+\frac{\delta}{3}+\frac{2\delta^2}{9}\right),\,\,\,
\]
\[
\delta=1-k,\,\,\,{\rm (around\,\,\, the\,\,\,
sphere)},\,\,|\delta|\ll 1.
\]
Solution of the system of equations (\ref{eq7})-(\ref{eq8p}) was
performed for initial conditions at $t=0$: $\dot c_0=0$, different
values of initial $a_0,\,\, \dot a_0,\,\,k_0$, and different
values of the constant parameter $\varepsilon$. Evidently, at
$k_0=1$, $\dot a_0=0$, $\varepsilon<1$ we have the spherical
collapse to singularity. The most interesting result was obtained
at $k_0 \neq 1$, and all other cases with deviations from
spherical symmetry. No singularity was reached in this case for
any $\varepsilon >0$. It is clear, that at  $\varepsilon =0$ a
weak singularity is reached during formation of a pancake with
infinite volume density, and finite gravitational force. At
$\varepsilon >0$ the behavior depends on the value of the total
energy $H$: at $H>0$ we obtain  a full disruption of the body, and
at $H<0$ the oscillatory regime is established at any value of
$\varepsilon<1$. At $\varepsilon\ge 1$ the total energy of
spheroid is positive, $H>0$, determining the full disruption. The
case with $H=0$ is described separately below.

At $H<0$ the type of oscillatory regime depends  on initial
conditions, and may be represented either by regular periodic
oscillations, or by chaotic behavior. Examples of two types of
such oscillations are represented in Figs.1-2 (regular, periodic),
and in Fig.3 (chaotic). For rigorous separation between these
kinds of oscillations we use a method developed by Poincar\'{e}
\citep{LL83}.

\begin{figure}
\centerline{\hbox{\includegraphics[width=0.5\textwidth]{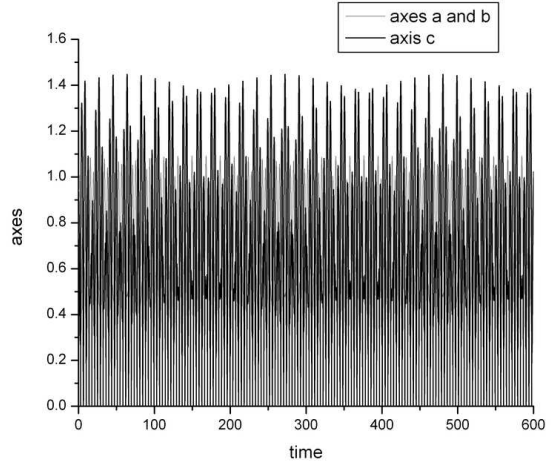}}}
\caption{Example of regular motion of spheroid with $\gamma=4/3$,
$H = - 1/5$, $\varepsilon=2/3$. This motion corresponds to full
line on the Poincar\'{e} map in Fig.4.}
\end{figure}

\begin{figure}
\centerline{\hbox{\includegraphics[width=0.5\textwidth]{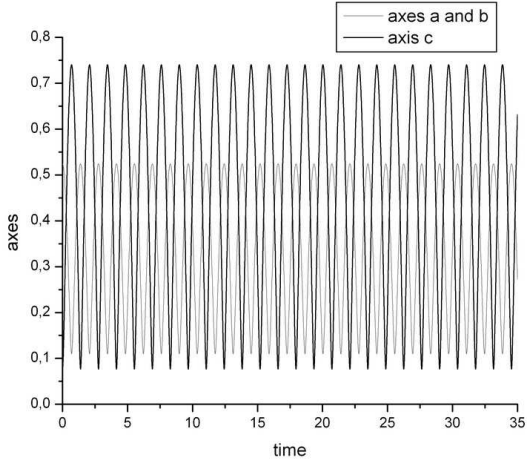}}}
\caption{Example of regular motion of spheroid with $\gamma=4/3$,
$H = - 1/5$, $\varepsilon=2/3$. This motion corresponds to the
point inside the regular region on the Poincar\'{e} map in Fig.4.}
\end{figure}

\begin{figure}
\centerline{\hbox{\includegraphics[width=0.5\textwidth]{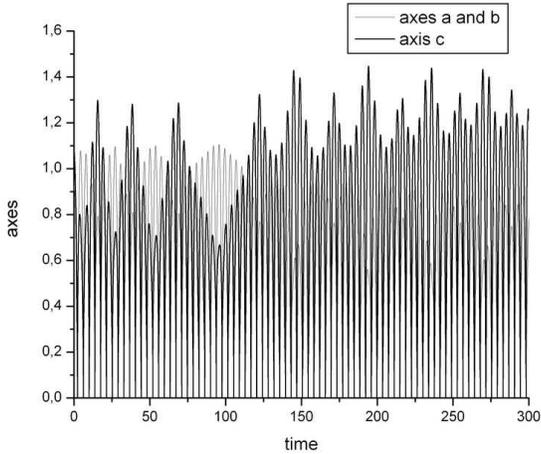}}}
\caption{Example of chaotic motion of spheroid with $\gamma=4/3$,
$H = - 1/5$, $\varepsilon=2/3$. This motion corresponds to gray
points on the Poincar\'{e} map in Fig.4.}
\end{figure}

\subsection{The case H=0}

In the case of a sphere  with zero initial radial velocity we have
equilibrium state with $\varepsilon = \varepsilon_{eq} = 1$. Let
us consider the case of $\dot{a} \neq 0$ and $\varepsilon < 1$. It
follows from (\ref{ham1}), that for non-zero  initial velocity
$\dot a$  entropy function $\varepsilon$ is strictly less than
unity. The fate of such sphere depends on the sigh of $\dot{a}$.
At $\dot{a}<0$ the star collapses to singularity, and at
$\dot{a}>0$ there total disruption happens, with zero velocity at
infinity.

For the initial spheroid with $a = b \neq c$, and zero initial
velocities $\dot{a} = \dot{c} =0$ the entropy function is also
less than 1. In this case the value of the entropy function
$\varepsilon$ is uniquely determined by the deformation $\delta =
(a-c)/a$. This dependence can be found explicitly from
(\ref{ham1}); for small $\delta$ we obtain

\begin{equation}
\varepsilon =  1 - \frac{4}{45} \, \delta^2.
\end{equation}
Thus even in the case of zero initial velocities, $\varepsilon$ is
less than unity for a spheroid, and reaches unity only for the
sphere. For the same $\varepsilon$ the deformation $\delta$ of the
zero energy body at rest have two states with $\delta=\pm
\frac{3}{2}\sqrt{5(1-\varepsilon)}$, corresponding to oblate and
prolate spheroids. If we set non-zero initial velocities in case of
spheroid, entropy function will be even less. Calculation of the
motion in this case leads finally to the expansion of the body, with
oscillating behavior of $\delta$ around zero (oblate - prolate
oscillations). It takes place both under the condition of the
initial contraction or initial expansion.

\section{The Poincar\'{e} section}

To investigate regular and chaotic dynamics we use the method of
Poincar\'{e} section \citep{LL83} and obtain the Poincar\'{e} map
for different values of the total energy $H$. Let us consider a
spheroid with semi-axes $a=b \neq c$. This system has two degrees
of freedom. Therefore in this case the phase space is
four-dimensional: $a$, $\dot{a}$, $c$, $\dot{c}$. If we choose a
value of the Hamiltonian $H_0$, we fix a three-dimensional energy
surface $H(a,\dot{a},c,\dot{c}) = H_0$. During the integration of
the equations (\ref{eq7})-(\ref{eq8p}) which preserve the constant
$H$, we fix moments $t_i$, when $\dot{c}=0$. At these moments
there are only two independent values (i.g.  $a$ and $\dot{a}$),
because the value of $c$ is determined uniquely from the relation
for the hamiltonian at constant
 $H$. At each moment $t_i$ we put a dot on the plane
$(a, \dot{a})$.

For the same values of $H$ and $\varepsilon$ we solve equations of
motion (\ref{eq7})-(\ref{eq8p}) at initial $\dot c=0$, and
different $a,\,\, \dot a$. For each integration we put the points
on the plane $(a, \dot{a})$ at the moments $t_i$. These points are
the intersection points of the trajectories on the
three-dimensional energy surface with a two-dimensional plane
$\dot{c}=0$, called the Poincar\'{e} section.

For each fixed combination of $\varepsilon, H$ we get the
Poincar\'{e} map, represented in Figs. 3-6. Condition $\dot{c}=0$
splits in two cases,  of a minimum and of a maximum of $c$. The
Poincar\'{e} maps are drawing separately, either for the minimum,
or for the maximum of $c$, and both maps lead to identical
results. The regular oscillations are represented by closed lines
on the Poincar\'{e} map, and chaotic behavior fills regions of
finite square with dots. These regions are separated from the
regions of the  regular oscillations by separatrix line.

\subsection{The bounding curve}

Actually, the variables $a$ and $\dot{a}$ cannot occupy the whole
plane
 $(a, \dot{a}): \; 0 < a < \infty, -\infty < \dot{a} <
+\infty$. Let us obtain a curve bounding the area of the values
$a$ and $\dot{a}$. Let a function $\Phi (a, \dot{a}, c)$ in the
variables (\ref{eqvar}), (\ref{eqvar1}) be

\[
\Phi (a, \dot{a}, c) = \frac{1}{10} \,
(\dot{a}^2+\dot{b}^2+\dot{c}^2) -
\]
\begin{equation}
-\frac{3}{10}\int\limits_0^{\infty}\frac{du}{\sqrt{(a^2+u)(b^2+u)(c^2+u)}}
+ \frac {3\varepsilon} {5(abc)^{1/3}} - H
\end{equation}
at $a=b$, $\dot{a}=\dot{b}$, $\dot{c}=0$, and fixed value of
$\varepsilon$ and $H$. The equation for the bounding curve
$f(a,\dot{a}) = 0$ at given $\varepsilon$ and $H$ is determined
from the following system of equation

\begin{equation}
\Phi (a, \dot{a}, c) = 0 \, , \quad \frac{\partial}{\partial c}
\Phi (a, \dot{a}, c) = 0 \, .
\end{equation}
 In case of $c<a$ this system of equations
has a form

\begin{equation}
\frac{1}{5} \, \dot{a}^2 - \frac{3}{5} \, \frac{\arccos ( c/a )}{
{ \sqrt {{a}^{2}-{c}^{2}}} } + \frac {3\varepsilon}{5(a^2
c)^{1/3}} - H = 0 \, ,
\end{equation}

\begin{equation}
-\frac{3}{5} \, c \, \frac{\arccos ( c/a )}{ ( {a}^{2}-{c}^{2}
)^{3/2} } + \frac{3}{5} \,  \frac{1}{a^2-c^2}  - \frac{1}{3} \,
\frac{1}{c} { \frac {3\varepsilon}{ 5({a}^{2}c ) ^{1/3}}} = 0 \, .
\end{equation}
This system is solved numerically. The second equation doesn't
depend on $\dot{a}$. We set $a$ and obtain corresponding value of
$c$ from the second equation. Then we substitute $a$ and $c$ into
the first equation and find $\dot{a}$. Thus we obtain the point
$(a, \dot{a})$. Changing $a$, we obtain numerically the curve
$f(a,\dot{a})=0$. This bounding curve is painted in Figs.4-8 by a
heavy line.

\begin{figure}
\centerline{\hbox{\includegraphics[width=0.5\textwidth]{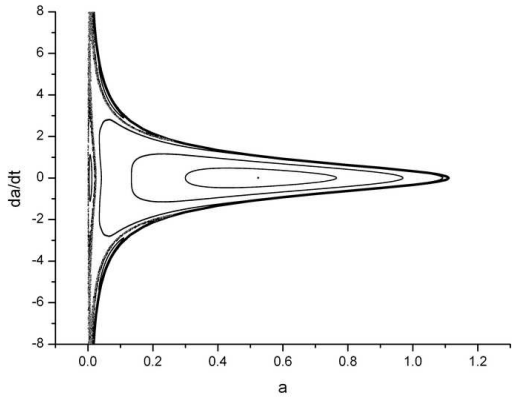}}}
\caption{The Poincar\'{e} map for five regular  and two chaotic
trajectories in case of $\gamma=4/3$, $H = - 1/5$,
$\varepsilon=2/3$. The $(a,\dot a)$ values are taken in the
minimum of c. Full black line is the bounding curve. The point
inside the regular region corresponds to coherent oscillations
with the same period for $a$ and $c$ values, represented in
Fig.2.}
\end{figure}

\begin{figure}
\centerline{\hbox{\includegraphics[width=0.5\textwidth]{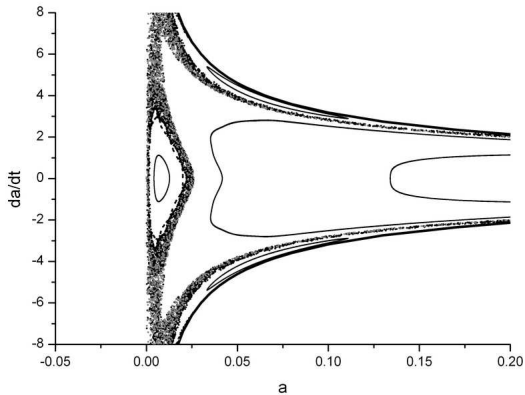}}}
\caption{Zoom of previous figure.}
\end{figure}

\begin{figure}
\centerline{\hbox{\includegraphics[width=0.5\textwidth]{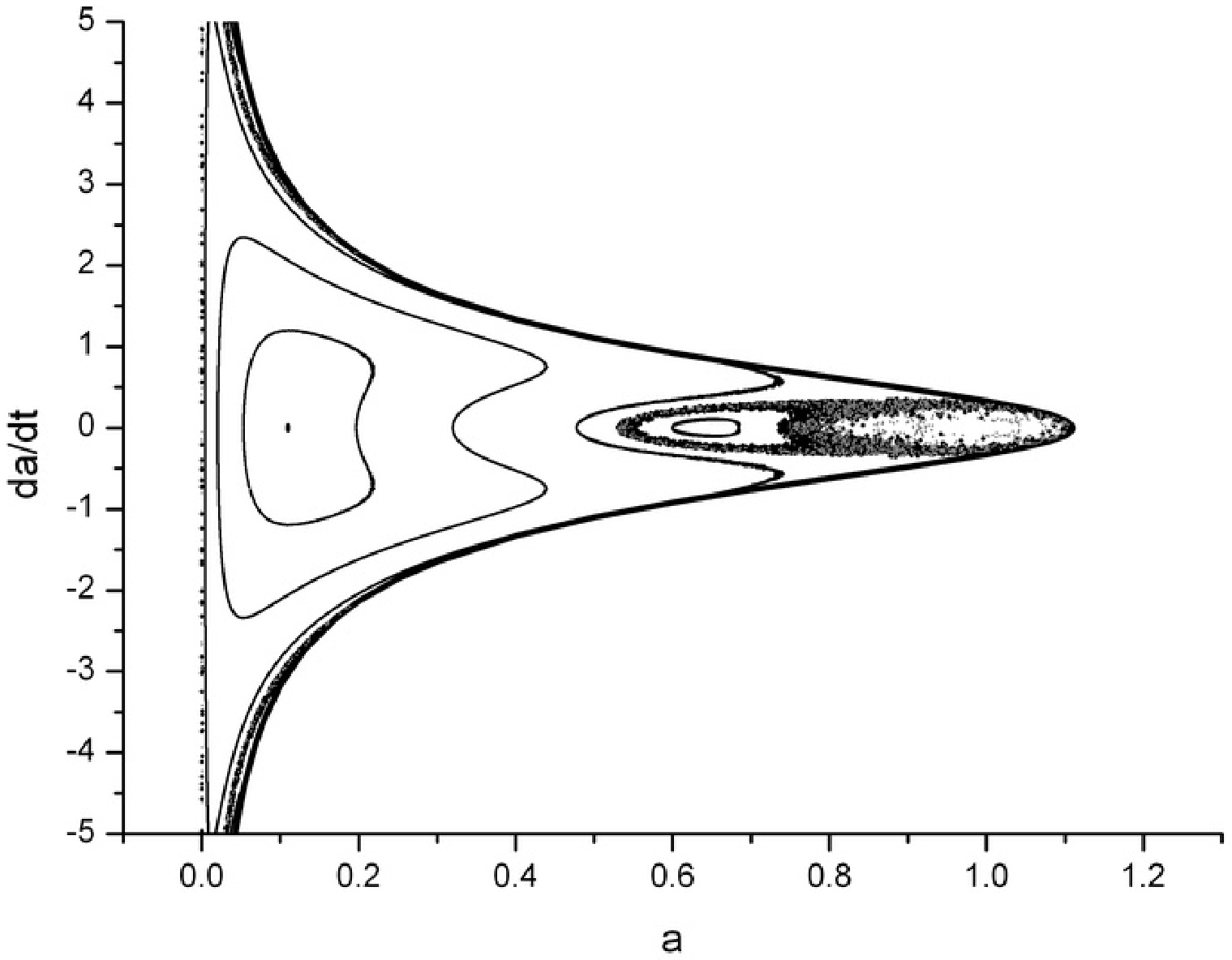}}}
%\centerline{\hbox{\includegraphics[width=1.2\textwidth]{Max1.eps}}}
\caption{The Poincar\'{e} map for five regular and two chaotic
trajectories in case of $\gamma=4/3$, $H = - 1/5$,
$\varepsilon=2/3$. The $(a,\dot a)$ values are taken in the
maximum of c. Full black line is the bounding curve. The point
inside the regular region corresponds to coherent oscillations
with the same period for $a$ and $c$ values, represented in
Fig.2.}
\end{figure}

%\begin{figure}
%\centerline{\hbox{\includegraphics[width=1.2\textwidth]{Max1zoom.eps}}}
%\caption{Zoom of previous figure.}
%\end{figure}

\begin{figure}
\centerline{\hbox{\includegraphics[width=0.5\textwidth]{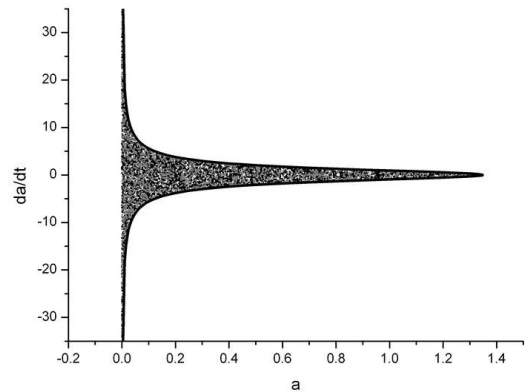}}}
\caption{The Poincar\'{e} map for two chaotic trajectories in case
of $\gamma=4/3$, $H = - 1/2$, $\varepsilon=1/6$. The $(a,\dot a)$
values are taken in the minimum of c. Full black line is the
bounding curve.}
\end{figure}

\begin{figure}
\centerline{\hbox{\includegraphics[width=0.5\textwidth]{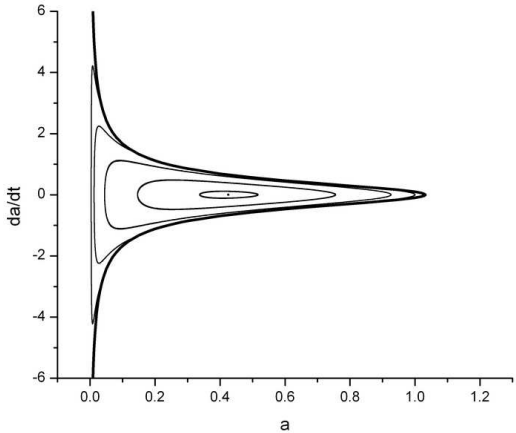}}}
\caption{The Poincar\'{e} map for six regular trajectories in the
case of $\gamma=4/3$, $H = - 3/50$, $\varepsilon=9/10$. The
$(a,\dot a)$ values are taken in the minimum of c. Full black line
is the bounding curve. The point inside the regular region
corresponds to coherent oscillations with the same period for $a$
and $c$ values, similar to those represented in Fig.2.}
\end{figure}

\section{Discussion}

The main result following from our calculations is the indication
to a degenerate nature of formation of a singularity  in unstable
newtonian self-gravitating gaseous bodies. Only pure spherical
models can collapse to singularity, but any kind of nonsphericity
leads to nonlinear stabilization of the collapse by a dynamic
motion, and formation of regularly or chaotically oscillating
body. This conclusion is valid for all unstable equations of
state, namely, for adiabatic with $\gamma <4/3$. In addition to
the case with $\gamma =4/3$, we have calculated the dynamics of
the model with $\gamma=6/5$, and have obtained similar results.
The Poincar\'{e} map for this case is represented in Fig.9. For
$\gamma=6/5$ we have the entropy function $\varepsilon = E_{th}
(abc)^{1/5}$, nondimensional entropy function $\tilde{\varepsilon}
= \varepsilon / U_0 a_0^{3/5}$, see (\ref{scal-par}). For the
spherical star with $\gamma=6/5$ the nondimensional Hamiltonian
instead of (\ref{sphere-H}) is
\begin{equation}
\label{gamma6} \tilde{H} = \frac{3m}{10} \dot{a}^2 - \frac{3}{5a}
m^2 + \frac{\tilde{\varepsilon}}{a^{3/5}} .
\end{equation}
Note, that region of chaotic behavior on the  Poincar\'{e} map is
gradually increasing for $\gamma=4/3$ with decreasing of the
entropy $\varepsilon$ and the total energy $H$. At
$\varepsilon=9/10$ and $H=-3/50$ we have found only regular
oscillations (Fig.8), at $\varepsilon=2/3$ and $H=-1/5$ both kind
of oscillations are present (Figs. 4-6), and only chaotic behavior
is found at $\varepsilon=1/6$ and $H=-1/2$ (Fig.7). We connect
\citep{BKTs2005} this chaotic behavior with development of
anisotropic instability, when radial velocities strongly exceed
the transversal ones \citep{ant, fp85}.

In reality a presence of dissipation leads to damping of these
oscillations, and to final collapse of nonrotating model, when
total energy of the body is negative. In the case of core-collapse
supernova the main dissipation is due to emission of neutrino. The
time of the neutrino losses is much larger than the characteristic
time of the collapse, so we may expect that the collapse leads to
formation of a neutron star where nonspherical modes are excited
and exist during several seconds after the collapse. In addition
to the damping due to neutrino emission the shock waves will be
generated, determining highly variable energy losses during the
oscillations. Besides viscosity and radiation which can damp the
ellipsoid-like motions and allow collapse, there is the
possibility that inertial interactions with higher order modes
that must be present in the real stratified bodies may cause an
inertial cascade and drain the energy from the second order modes
an a non-secular way that does not depend on dissipative
coefficients. It is very seductive to connect chaotic oscillations
with highly variable emission observed in the prompt gamma ray
emission of cosmic gamma ray bursts \citep{mg81}. A presence of
rotation and magnetic field strongly complicate the picture of the
core-collapse supernova explosion \citep{abkm05}.

In this paper we consider in details only spheroidal bodies. In
reality the spheroid will become a triaxial ellipsoid during the
motion. In addition to spheroids we calculate many variants with
triaxial figures (see also \citet{BKTs2005}). Qualitatively we
obtain the same results for ellipsoids: no singularity was reached
for any $\varepsilon > 0$ and establishing oscillatory (regular or
chaotic) regime under negative total energy prevents the collapse.
However, in the case of the ellipsoid with semi-axes $a \neq b
\neq c$ we have a system with three degrees of freedom and
six-dimensional phase space. Therefore we could not carry out
rigorous investigation of regular and chaotic types of motion by
the constructing Poincar\'{e} map as it was done for spheroid with
two degrees of freedom, and restricted ourselfs by a description
of the spheroidal case.

In the frame of a general relativity dynamic stabilization against
collapse by nonlinear nonspherical oscillations cannot be
universal. When the size of the body approaches gravitational
radius no stabilization is possible at any $\gamma$. Nevertheless,
the nonlinear stabilization may happen at larger radii, so after
damping of the oscillations the star would collapse to the black
hole. Due to development of nonspherical oscillations there is a
possibility for emission of gravitational waves during the
collapse of nonrotating stars with the intensity similar to
rotating bodies, or even larger.

Account of general relativity will introduce a new non-dimensional
parameter, which can be written as $p_g = \frac{2 G m_0}{c^2
a_0}$. The fate of the gravitating body will depend on the value
of this parameter, and we may expect a direct relativistic
collapse to a black hole at increasing $p_g$, approaching unity.
It is known, that a nonrotating black hole is characterized only
by its mass \citep{ZN}. In absence of other dissipative processes,
the excess of energy, connected with a nonspherical motion should
be emitted by gravitational waves during a formation of the black
hole.

\begin{figure}
\centerline{\hbox{\includegraphics[width=0.5\textwidth]{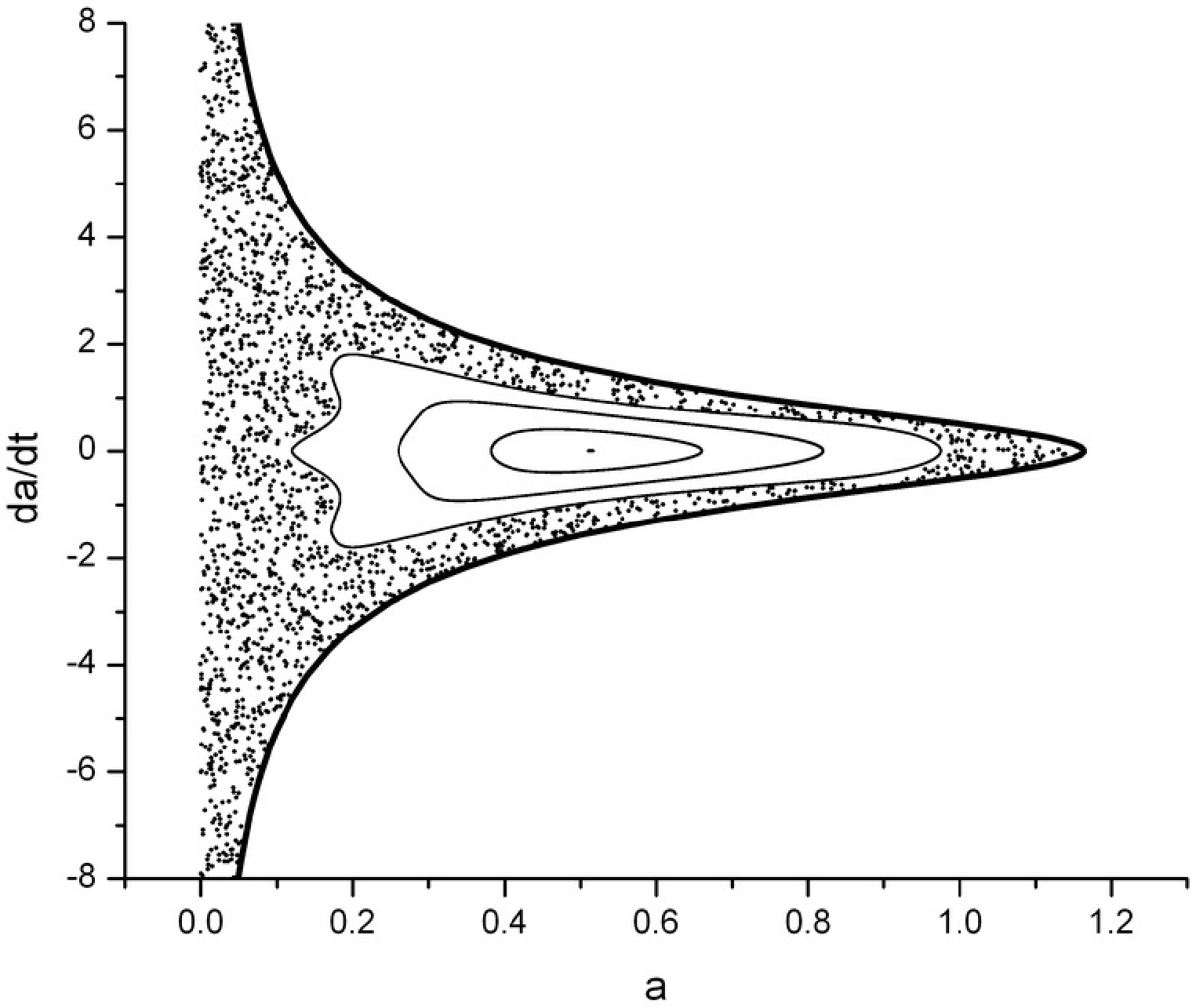}}}
\caption{The Poincar\'{e} map for one chaotic and four regular
trajectories in case of $\gamma=6/5$, $m=1$, $\tilde H = - 3/50$,
$\tilde\varepsilon=27/50$, see (\ref{gamma6}). The $(a,\dot a)$
values are taken in the minimum of c. Full black line is the
bounding curve. The point inside the regular region corresponds to
coherent oscillations with the same period for $a$ and $c$ values,
similar to those represented in Fig.2.}
\end{figure}

\section*{Acknowledgments}
We are grateful to A.I. Neishtadt for the valuable advices in
constructing the Poincar\'{e} map, and to anonymous referee for
useful remarks.

This work was partially supported by RFBR grants 05-02-17697,
06-02-90864 and 06-02-91157, RAN Program "Formation and evolution
of stars and galaxies" and Grant for Leading Scientific Schools
NSh-10181.2006.2.

\bsp

\label{lastpage}

\end{document}